\documentclass[review]{elsarticle}

\usepackage{lineno,hyperref}
\usepackage{xspace}
\modulolinenumbers[200]
\usepackage{bm}
\usepackage{amssymb}
\usepackage{hyperref}
\usepackage{amsmath}
\usepackage{csquotes}
\usepackage[dvipsnames]{xcolor}

\usepackage{lineno,xcolor}
\linenumbers
\journal{Journal of \LaTeX\ Templates}

\begin{document}

\begin{frontmatter}

\title{Parton energy loss in pp collisions at very high multiplicity}

\author{Aditya Nath Mishra and Guy Pai{\'c}}
\address{Instituto de Ciencias Nucleares, Universidad Nacional Aut\'onoma de M\'exico, \\ Apartado Postal 70-543,
	M\'exico City 04510, M\'exico}

\begin{abstract}

We present the results for the evolution of transverse momentum spectra for the jetty and underlying parts of events as a function of multiplicity in pp collisions at 13 TeV measured at midrapidity ($\left|\eta\right|<0.8$) using PYTHIA8 event generator. The main characteristic of the approach is that it reaches to extreme multiplicities not yet explored by the experiments. We demonstrate that the behavior of both the underlying and hard components are affected by the multiplicity of the events i.e. the energy density. The behavior of the spectra at very high multiplicities suggests that the partons suffer energy loss compensated by an increase in the multiplicity of events.  

\end{abstract}

\begin{keyword}
pp collisions, underlying events, multiplicity dependence, leading particles 
\end{keyword}

\end{frontmatter}

The high multiplicity events in pp collisions have spurred quite an interest in the last few years for the similarities they show with heavy-ion collisions~\cite{ALICE:2011ac,Ortiz:2017jaz,Armesto:2015kwa,Alvioli:2016prc,Ortiz:2019prd,Loizides:2016tew}. The similarities were acknowledged but with a caveat: namely, no experimental analysis could give evidence for the \enquote{jet quenching} in pp collisions. We present the results of a simulation of PYTHIA8 which demonstrates that, at the highest achieved multiplicities, hence the highest energy densities; we observe important modifications of the particle momentum spectra and a simultaneous increase of the yield of the Underlying Event (UE).  To our knowledge, we do not have, yet, experimental proof of a similar mechanism. The highest deliberate attempt at high multiplicity has been done by the LHC experiments resulting with the interesting \enquote{ridge} discovery~\cite{CMS:2010ridge}. However, the largest multiplicity dealt with, reported in ref~\cite{CMS:2013epjc}, reached only up to 140 charged particles in a pseudorapidity range of $|\eta| < 2.4$.
In the present study, we reach much higher multiplicities, up to more than 100 charged particles, in a very central pseudorapidity range of $|\eta| < 0.8$.

\begin{figure}[ht!]
	\begin{center}
	\includegraphics[keepaspectratio, width=1.0\columnwidth]{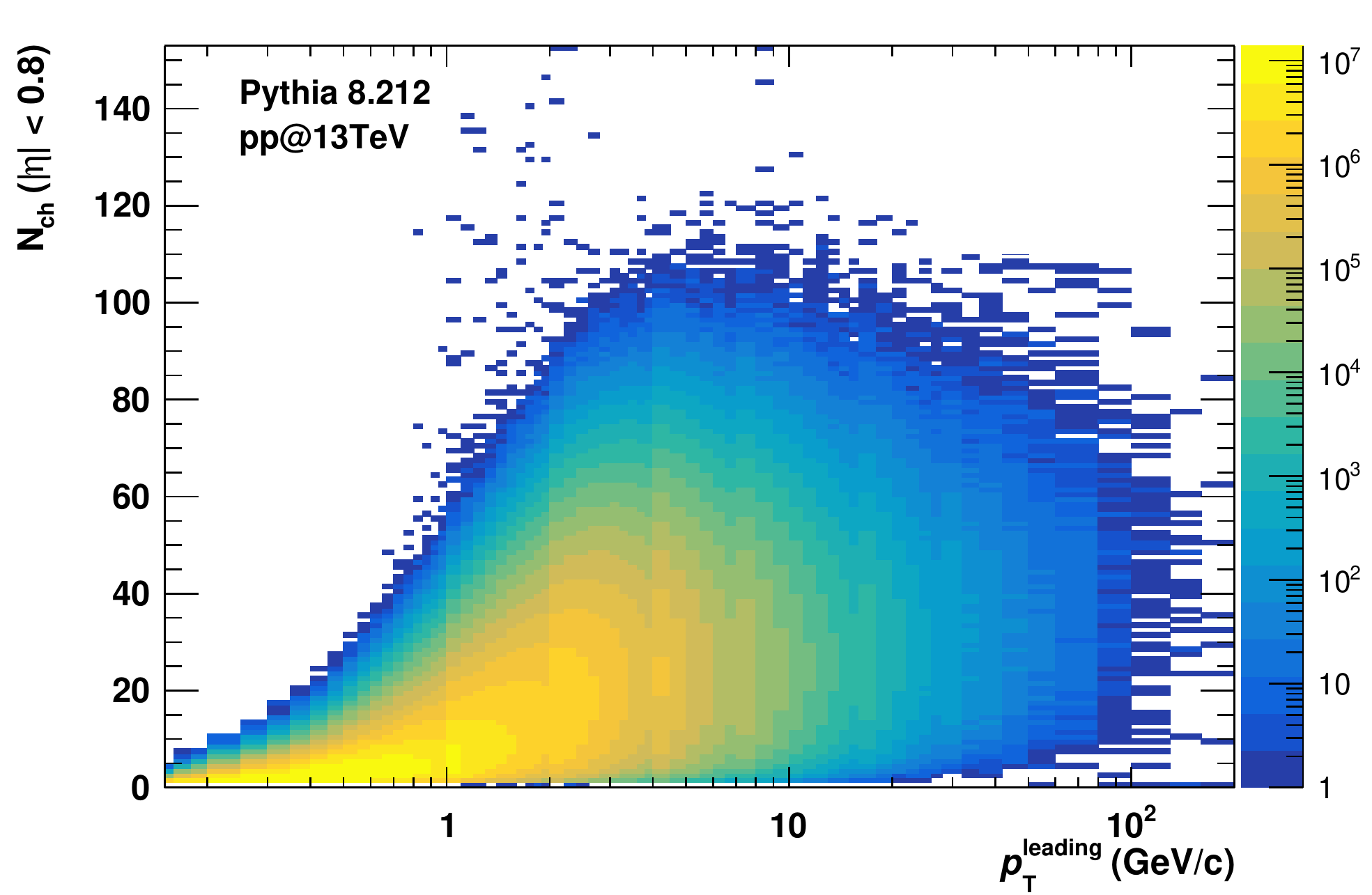}
		\caption{\label{fig:0} Correlation plot between charged particle multiplicity and leading charged particle transverse momentum of the event at mid-rapidity.}
	\end{center}
\end{figure}

\begin{figure*}[ht!]
	\begin{center}
		\includegraphics[keepaspectratio, width=1.0\columnwidth]{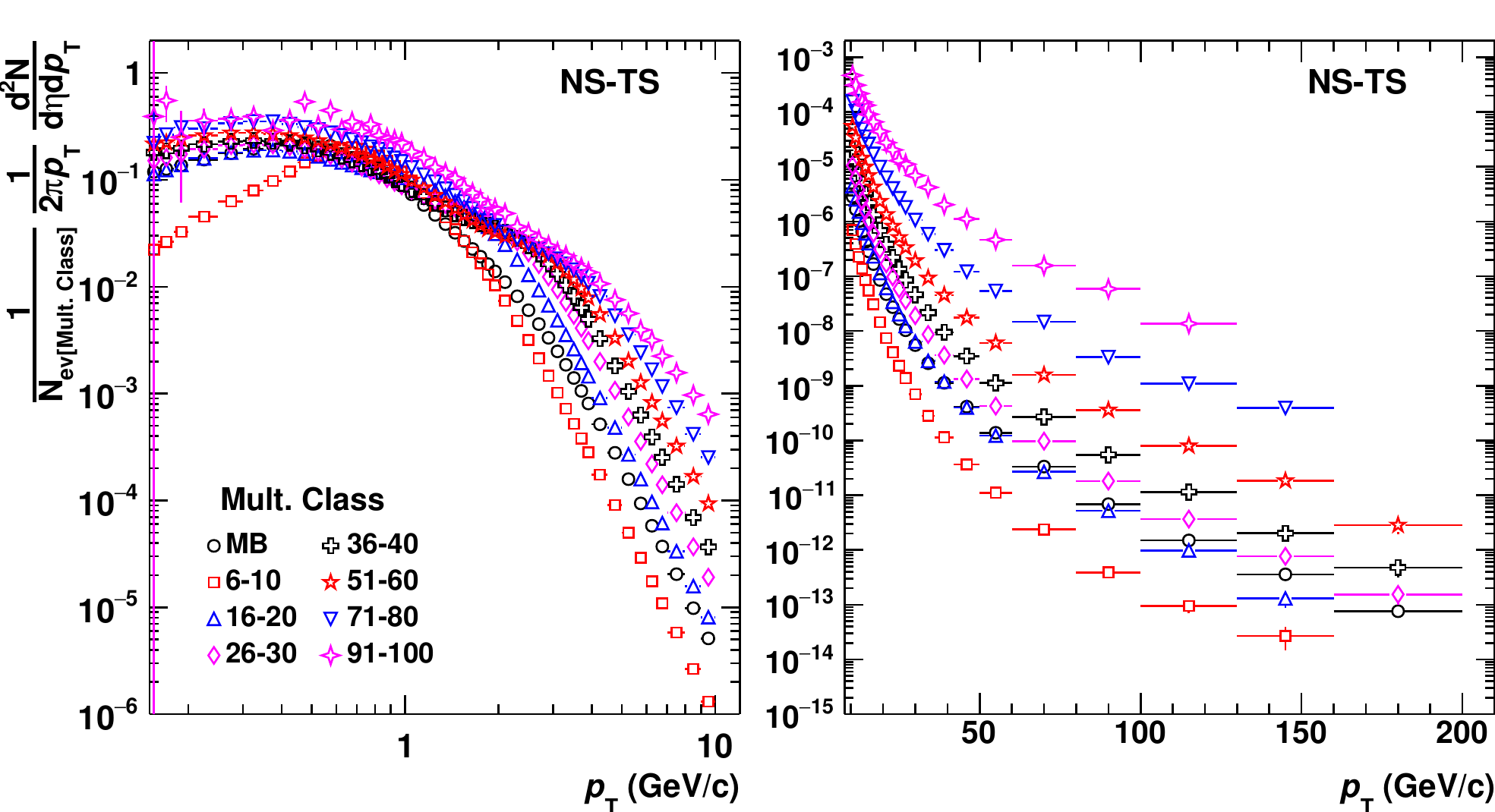}
		\caption{\label{fig:1}Transverse momentum spectra of charged particles in function of multiplicity  obtained subtracting the contribution of TS spectra from the NS region ones. The figure is split for better lisibility into two panels for the low $\ensuremath{p_{\rm{T}}}$ region (left) and the high $\ensuremath{p_{\rm{T}}}$ ones (right).} 
	\end{center}
\end{figure*}

\begin{figure*}[t!]
	\begin{center}
		\includegraphics[keepaspectratio, width=1.0\columnwidth]{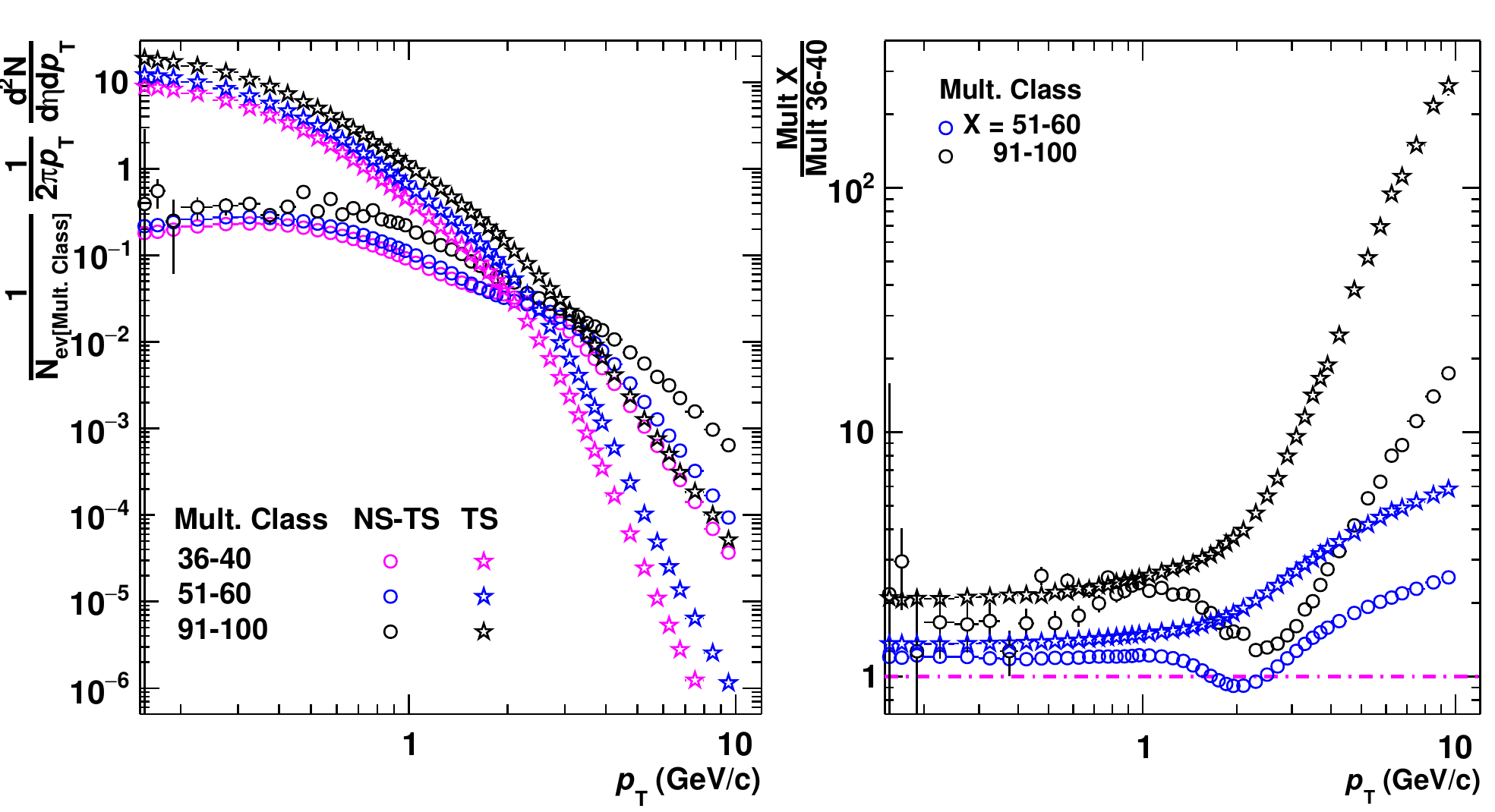}
		\caption{\label{fig:2}(Left) The NS-TS spectra compared with the TS ones for the low-$\ensuremath{p_{\rm{T}}}$ part of spectra (up to 10 GeV/$c$) for 3 multiplicities (see text). (Right) the ratios of the two highest multiplicity spectra with the lowest multiplicity one.} 
	\end{center}	
\end{figure*}

\begin{figure*}[t!]
	\begin{center}
		\includegraphics[keepaspectratio, width=1.0\columnwidth]{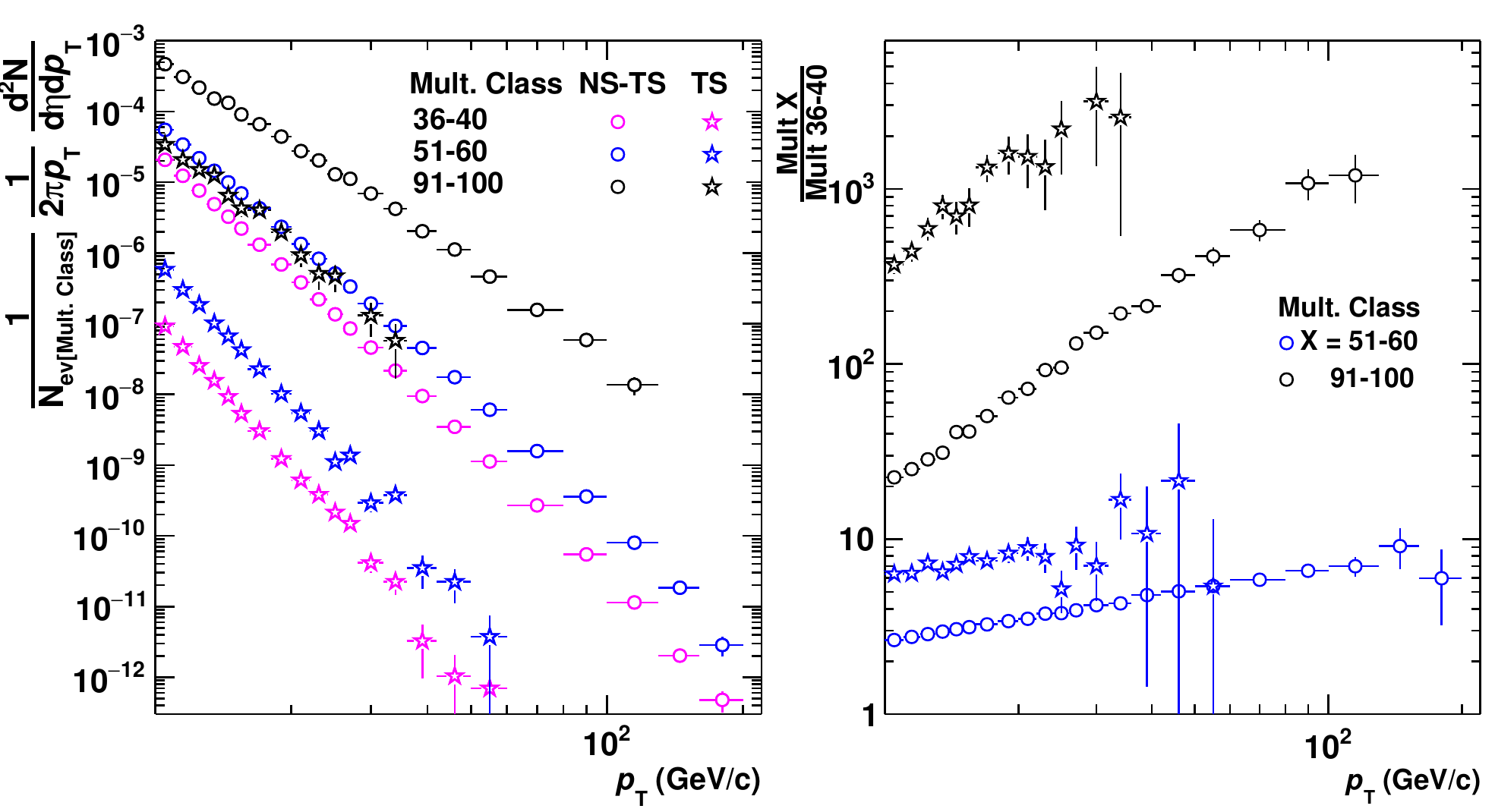}
		\caption{\label{fig:3}(Left) The NS-TS spectra compared with the TS ones for the high-$\ensuremath{p_{\rm{T}}}$ part of spectra (from 10 to 200 GeV/$c$) for 3 multiplicities. (Right) the ratios of the two highest multiplicity spectra with the lowest multiplicity one.} 
	\end{center}	
\end{figure*}

Originally, our interest was provoked by a correlation plot  between the jet energy and event multiplicity  in the thesis of Benjamin Hess~\cite{HESS:2015thesis} (in unpublished ALICE data), where one observes that, in the two-dimensional plot, multiplicity vs. jet transverse momentum for 7 TeV collisions of protons, the maximum jet energy is obtained at relatively low multiplicity. This is in contrast with the well accepted observation that with multiplicity ever higher momentum particles are produced. A similar observation has been reported by Zolt{\'a}n Varga {\it et al.} in ref~\cite{Varga:2019ahep} using the PYTHIA event generator. 


Since the PYTHIA MC generator provides generally satisfactory reproduction of the pp collisions at LHC, we decided to investigate the correlations of the leading transverse momentum and the  inclusive charged particle spectra in  different azimuthal regions defined by the leading particle, with multiplicity till the highest possible values. We believe that it is legitimate to study
these correlations
at the extremes, to bring light to the behavior of partons in a very dense environment created in pp collision~\cite{Silvia:2018}. We present the results obtained with the simulation of $2{\rm x} 10^{9} $ events using the PYTHIA 8.212~\cite{pythia:2015} (Monash 2013 tune~\cite{Monash:2014epjc}) within the $|\eta| < 0.8$ acceptance for pp events at $\sqrt {s} $ = 13 TeV.

As a first check, a 2-Dimensional plot for the charged particle multiplicity versus transverse momentum of the leading charged particle of the event ($p_{\rm T}^{\rm leading}$) at mid-rapidity  ($|\eta|< 0.8$) is shown in Fig.~\ref{fig:0}. Note that we use threshold $p_{\rm T} > 0.15~GeV/c$ to keep the environment similar to the one at A Large Ion Collider Experiment (ALICE) at LHC. From the plot it can be seen that the highest  transverse momentum particles are produced for events with multiplicities of 30-60, while the highest multiplicity attainable in this analysis is twice as large. 
The conventional view~\cite{RickField:2011,RickField:2012} of producing the leading particle is based on the hard scattering of partons, with the accompanying  particles due to various mechanisms (initial and final state radiation, beam remnants etc).  This picture has brought a very successful industry of analysis of the so called Underlying Event (UE) and its correlation to the leading particle momentum. In the present study we are challenging this approach pointing to  alternative processes that seem to be present. In the following, we give a brief account of the technique to extract the  underlying  event structure. 

\begin{figure*}[t!]
	\begin{center}
		\includegraphics[keepaspectratio, width=1.0\columnwidth]{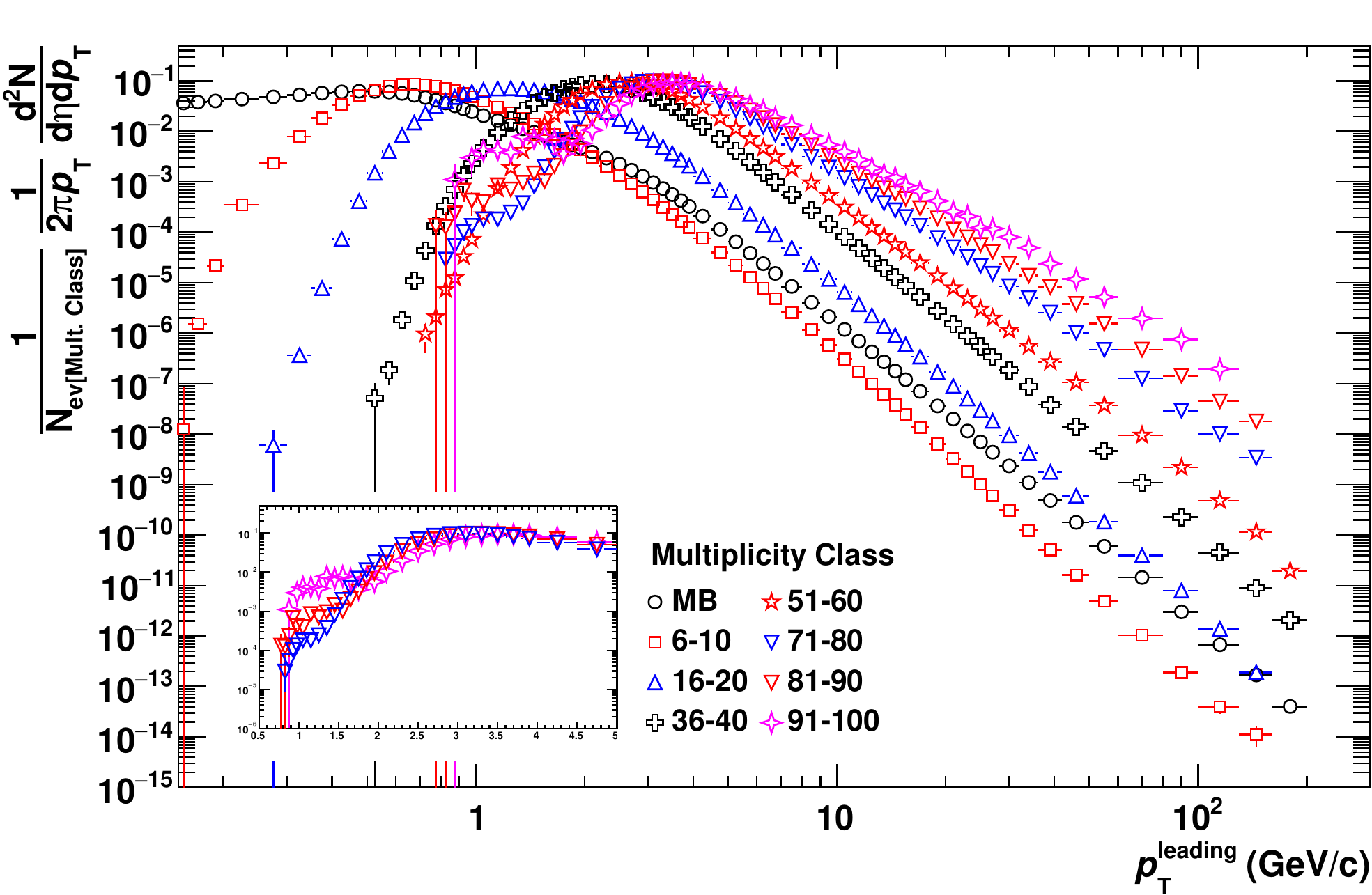}
		\caption{\label{fig:4} Leading particle spectra of charged particles for minimum bias and different multiplicity classes. Inset$:$ Low-$\ensuremath{p_{\rm{T}}}$ part of the spectra for the three highest multiplicities in the linear $\ensuremath{p_{\rm{T}}}$ scale.} 
	\end{center}	
\end{figure*}

\begin{figure}[h!]
	\begin{center}
		\includegraphics[keepaspectratio, width=1.0\columnwidth]{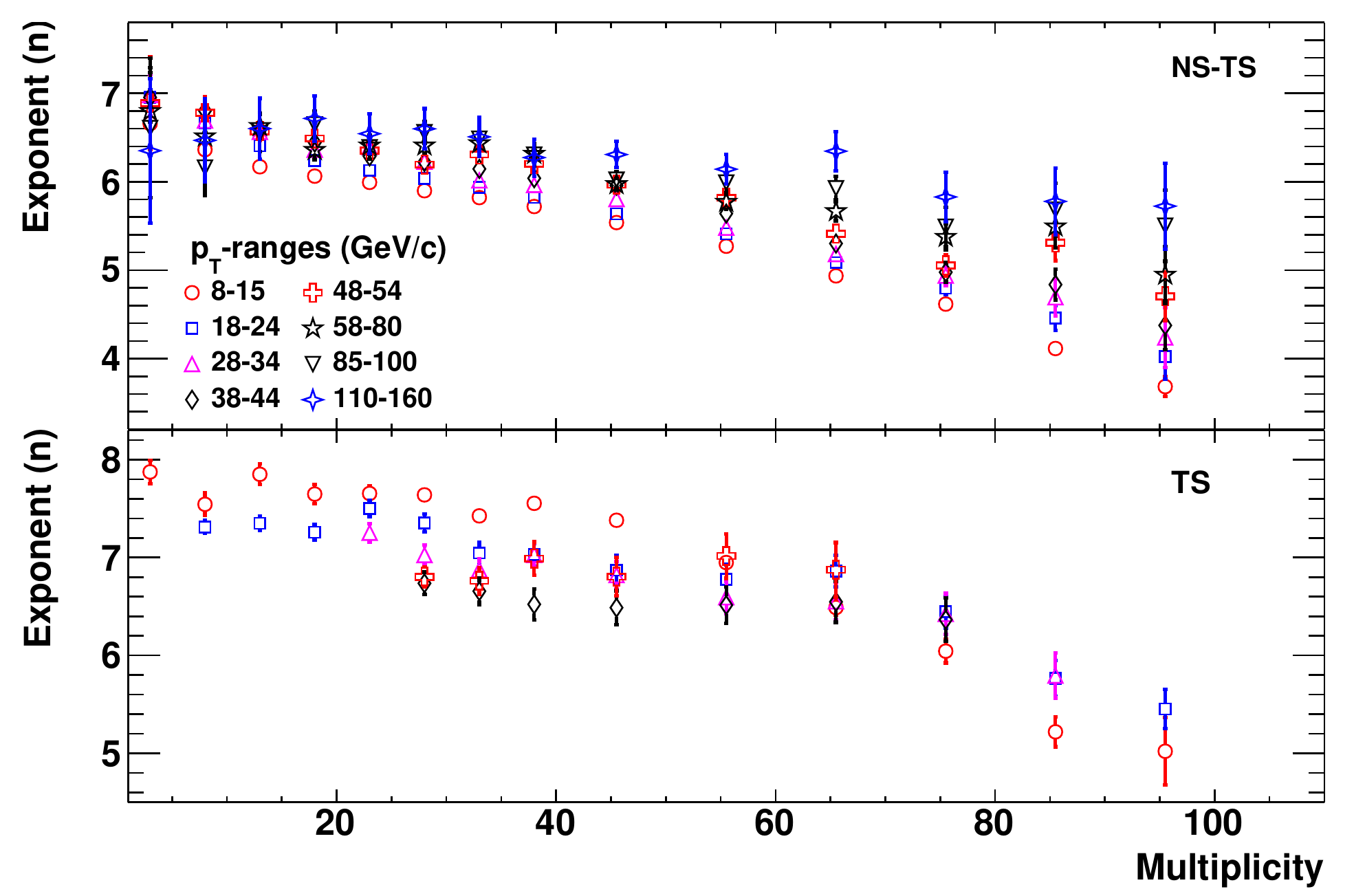}
		\caption{\label{fig:5}Distribution of power-law exponents obtained fitting the NS-TS (top) and TS (bottom) spectra  in different $\ensuremath{p_{\rm{T}}}$ regions in function of the multiplicity  of the events. The $\ensuremath{p_{\rm{T}}}$ ranges indicate the bins used to calculate the exponents. 
		} 
	\end{center}	
\end{figure}

\begin{figure}[h!]
	\begin{center}
		\includegraphics[keepaspectratio, width=1.0\columnwidth]{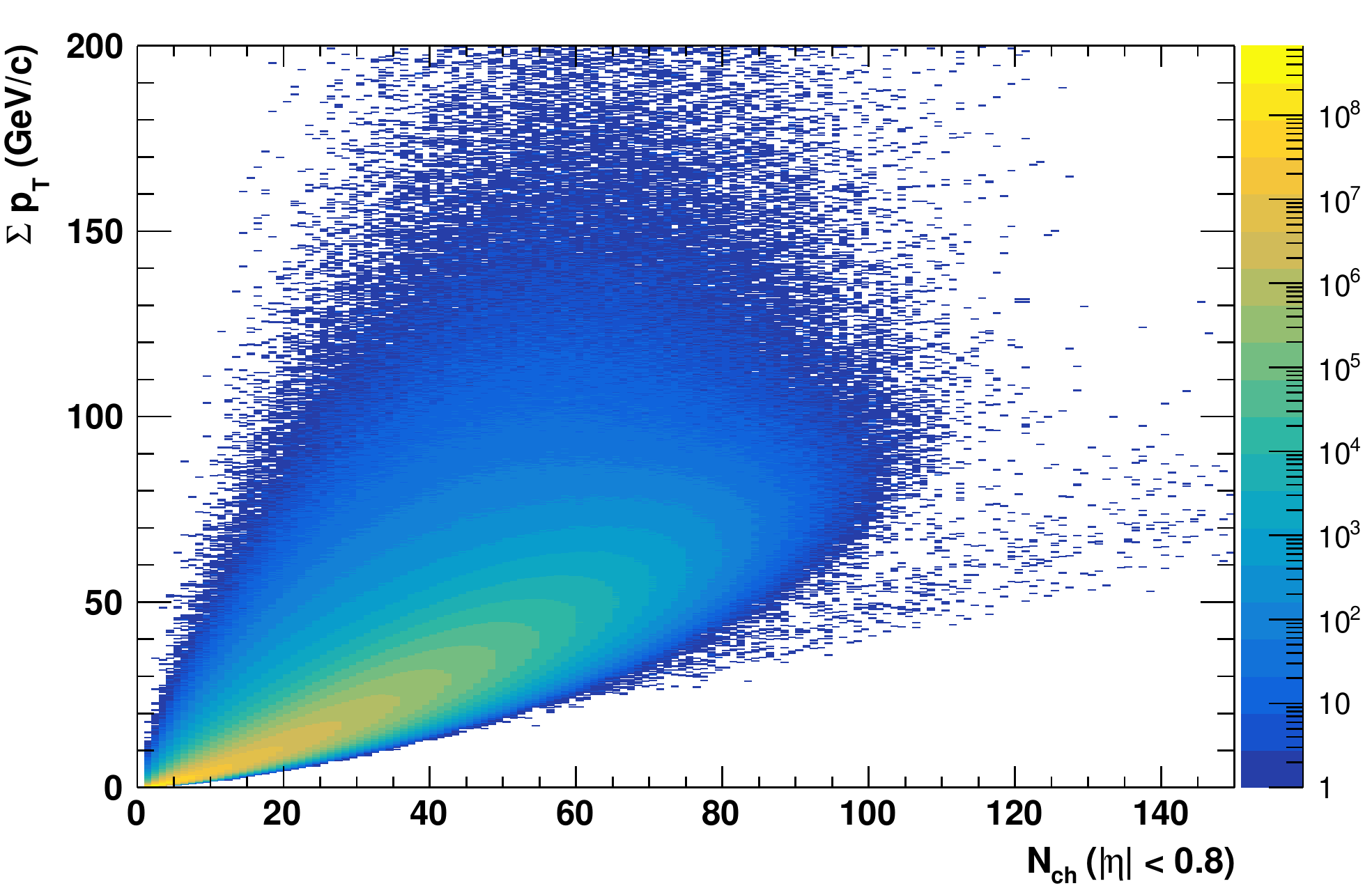}
		\caption{\label{fig:sumpt} The behavior of the sum of transverse momenta in function of multiplicity Distribution.
		} 
	\end{center}	
\end{figure}

We have adopted the CDF definition of the UE. The CDF definition splits the  azimuthal region of emission with respect to the  leading particle into 4 regions as follows: the  so called Near Side (NS) region defined by an angle of 60 degrees on each side of the leading particle \footnote{In the present analysis the leading particles are included in Near Side spectra.}, while the Away Side (AS)  region is defined with the same angular opening around the direction 180 degrees opposite to NS. Finally, on each side we have 2 regions covering the so called Transverse Side (TS) region~\cite{Affolder:2002prd}. In the present article we limit ourselves to the description of the results considering only the NS and TS regions.
The TS region has been extensively used to investigate the UE and its dependence on the leading particle momentum. The usual approach is to study the yields of charged particles in the TS region in function of the leading particle transverse momentum observed in the NS region. The main conclusion is that the UE has a strong initial rise until a leading particle $\ensuremath{p_{\rm{T}}}$ of $\approx$ 5 GeV/$c$~\cite{RickField:2011,RickField:2012} followed with a slower rise in function of the leading $\ensuremath{p_{\rm{T}}}$, while the NS continues rising at a higher rate suggesting that the correlation with the TS is not strong. Unfortunately, the studies published so far treat only the minimum bias data in pp collisions. The minimum bias measurements forcibly are dominated by low multiplicity events and therefore are occulting the high multiplicity ones, which correspond to high energy densities. Our approach derives from the knowledge that many features emerge only in high multiplicity pp events~\cite{mishra:2019prc,Siga:2019}. We therefore investigate the behavior of spectra in the NS and TS regions in function of the multiplicity. Similar, but not identical approaches were reported by the  CDF collaboration which studied the TS region $\ensuremath{p_{\rm{T}}}$ spectra in function of several leading jet energies in the NS~\cite{RickField:2011,RickField:2012}.
We have introduced a spectrum labeled NS-TS which is obtained by subtracting the TS spectrum from the NS one since the TS one is supposed to run also in the NS region. The resulting spectrum represents thus \enquote{hard/jetty} components. The results are reported in Figs.~\ref{fig:1},~\ref{fig:2} and~\ref{fig:3}. 


In Fig.~\ref{fig:1}, we present the evolution of the spectra  NS-TS with multiplicity in the central pseudorapidity region. In Fig~\ref{fig:2} and~\ref{fig:3}, we compare the behavior of the TS spectra for 3 multiplicities (not to obstruct the figure with too many multiplicities) with the NS-TS ones. We chose multiplicities below, at, and above the one giving the maximum $\ensuremath{p_{\rm{T}}}$ reach. The spectra plotted in Figs.~\ref{fig:1},~\ref{fig:2} and~\ref{fig:3} are normalized to the number of events in every multiplicity bin. 
The figures invite the following observation:
\begin{enumerate}
	\item For all multiplicities, the spectra in the both regions (NS-TS and TS) exhibit a hardening with multiplicity
	\item The maximum leading particle transverse momenta in the present   analysis are reached for multiplicities  of $\approx$ 50 charged particles, while {\bf at higher multiplicities the  slopes of the spectra continuously decreases without higher momentum particles!} Actually, one observes that at densities below  $\approx$ 50 the slopes of all the multiplicity bins are approximately equal while above the  critical charged particle density the slopes get gradually smaller both for the NS-TS and TS spectra. 
	We document this behavior in the right parts of Figs.~\ref{fig:2} and~\ref{fig:3} where we plot the ratios of the spectra of the two high multiplicity spectra to the low multiplicity one. 
	\item One observes, both in the NS-TS and TS spectra, a very significant and rapid increase in the ratios with the transverse momentum. The observed increase is much greater for the highest multiplicity than for the lower one by more than an order of magnitude at 100 GeV/$c$! The minima in the ratios for the NS-TS spectra are due to the position of the maxima of the leading particle spectra (see  Fig.~\ref{fig:4}).
	\item The TS spectra show an even larger ratio with the transverse momentum as visible from Figs.~\ref{fig:2} and~\ref{fig:3}.  
	\item At high multiplicities ($>$70), the leading charged particles spectra seem to have a \enquote{kink} at low-$\ensuremath{p_{\rm{T}}}$ (below 1 GeV/$c$) which is not visible at lower multiplicities! (see Fig.~\ref{fig:4})
\end{enumerate}

The general conclusion is that the two regions have an important correlation in function of multiplicity. The results may be summarized as follows: while at moderate multiplicities, we observe a gradual increase in the maximum $\ensuremath{p_{\rm{T}}}$ of the events in NS-TS accompanied  with a hardening in the TS region. At multiplicities above $\approx$ 50, the production of the highest momentum particles decreases while the mean transverse momentum in the TS continuously rises! 
The observations above invite an important question: are we observing some kind of energy loss of the highest momentum partons $-$ producing particles at lower momenta, increasing thus the multiplicity and mean momentum?

The production of jets is usually described as the  product of 3 fundamental quantities: the parton distribution function, the cross-section for a hard process and finally the fragmentation function which describes the probability for the outgoing parton to fragment into final hadrons.  None of the above mentioned factors seem a candidate to reduce the production of high $\ensuremath{p_{\rm{T}}}$ particles at high multiplicities.  However, the simulation does show:  both, a decrease in the maximum leading particle momentum, and an increase in the mean momentum of the particle spectra.

Additionally, in Fig.~\ref{fig:4}, we plot the spectra of leading particle transverse momenta in function of multiplicity. We observe that for the highest multiplicity bins the low $\ensuremath{p_{\rm{T}}}$ part of the spectra develop a \enquote{kink} at around 1 GeV/$c$ suggesting that the leading particles have been \enquote{degraded}.

In Fig.~\ref{fig:5}, we demonstrate this evolution by depicting the behavior of the power-law exponents, obtained by fitting the spectra in different $\ensuremath{p_{\rm{T}}}$ ranges, from the low $\ensuremath{p_{\rm{T}}}$ range 8-15 GeV/$c$  to the high momentum one 85-100 GeV/$c$ for different multiplicity ranges.  We observe a rather important variation in the power-law exponent in the low $\ensuremath{p_{\rm{T}}}$ region in the manner of Ref~\cite{mishra:2019prc}, beyond the multiplicities corresponding to the maximum leading transverse momenta, while in the higher $\ensuremath{p_{\rm{T}}}$ bins this tendency is much smaller, and actually the values are close to the minimum bias ones.  

Finally, we present in Fig.~\ref{fig:sumpt}, the behavior of the sum of transverse momenta in function of multiplicity in a 2D plot. The plot reveals another important feature: the maximum value  IS NOT achieved for the Highest multiplicities! This observation indicates a change in the hadrochemistry in high multiplicity events a feature that has been also observed experimentally~\cite{ALICE:2019Mult,Trzeciak:MPI2019}. We believe that what is usually claimed to be  Mass ordering  of the heavier masses is in fact the added production of massive particles. Unfortunately the limited PID capabilities do not permit  observations in a larger momentum range. 

\section*{Conclusions}
We have presented the results of simulations in the evolution of the leading particle spectra and the Underlying Event in a novel manner i.e. analyzing the prediction of the  generator Pythia for the momentum distributions in function of the multiplicity of the proton proton collisions at 13 TeV.  
The study has revealed the following unorthodox features:
\begin{enumerate}
	\item The maximum reachable multiplicities are not accompanied by an increase in the maximum leading particle momentum. The proportionality between maximum $\ensuremath{p_{\rm{T}}}$ and increasing multiplicity breaks down at multiplicities of 30-60 charged particles in a very central pseudorapidity range of $|\eta| < 0.8$\footnote{PYTHIA simulations made for larger $\eta$ ranges  show essentially the same behavior}.
	
	\item Beyond that multiplicity density the NS-TS spectra continue to get flatter, increasing the mean transverse momentum, seemingly, at the expense of the maximum reachable momentum.   
	\item Beyond the  particle density corresponding to the maximum $\ensuremath{p_{\rm{T}}}$ reach both the TS and the NS-TS spectra suffer a sudden hardening as visible in Figs.~\ref{fig:2},~\ref{fig:3} and~\ref{fig:5}. 
	
	\item At very low momenta the high multiplicity events present also a specific evolution by augmenting the yield at the smallest transverse momenta. The feature is observed both in the NS-TS and in the leading particle spectra. A similar behavior has been observed in ALICE~\cite{Siga:2019}.
\end{enumerate}

We infer from these observations that, in some way,  when very high energy densities are produced  a loss of the particle momenta is occurring, increasing the multiplicity and degrading the momenta of the particles. We should bear in mind that the maximum multiplicities investigated here do belong to events where there is a an important overlap of hadrons in a very small volume~\cite{Silvia:2018}. With the run 3 planned at the LHC one could open a very interesting and important area. Finally, it is interesting that the present results,  seen in the low momenta only, would be perfectly compatible with the hydrodynamic picture of the evolution of the collisions. To disentangle both alternatives it is imperative measure spectra to high $\ensuremath{p_{\rm{T}}}$ and at very high multiplicity.

\section*{Acknowledgements}
We acknowledge the discussions and suggestions of A. Ortiz throughout the development of the work. G.P. thanks the DGAPA, the Centro Fermi for their support. A.M. acknowledges the post-doctoral fellowship of DGAPA UNAM.

\section*{References}
\bibliography{MyRefFile}
\end{document}